\documentclass[multi]{cambridge7A}
\usepackage[UKenglish]{babel}
\usepackage[longnamesfirst,sectionbib]{natbib}
\usepackage{chapterbib}
\usepackage{amsmath,amssymb,amsthm,amsfonts}
\usepackage{enumerate,url}
\usepackage{graphicx}
\usepackage{color}

\setcitestyle{numbers,square,comma}
\allowdisplaybreaks[1]

\theoremstyle{plain}
\newtheorem{theorem}{Theorem}[section]
\newtheorem{lemma}[theorem]{Lemma}

\providecommand*\Index[1]{#1\index{#1}}
\providecommand*\undex[1]{} 
\providecommand*\Undex[1]{#1} 

\def\ignore#1{{}}

\newcommand*{\bigtimes}{\mathop{\raisebox{-.5ex}{\hbox{\huge{$\times$}}}}}

\newcommand{\E}{\mbox{\bf E}}
\newcommand{\pr}{\mbox{\bf P}}
\newcommand{\m}[1]{\marginpar{\tiny{#1}}}

\newcommand{\ints}{\mathbb{Z}}

\newcommand{\dtv}{d_{\mbox{\tiny TV}}}
\def\integ{\ints}
\def\ex{\E}

\newcommand{\eqa}{\begin{eqnarray}}
\newcommand{\ena}{\end{eqnarray}}
\newcommand{\eq}{\begin{equation}}
\newcommand{\en}{\end{equation}}
\newcommand{\eqs}{\begin{eqnarray*}}
\newcommand{\ens}{\end{eqnarray*}}

\def\l{\lambda}

\def\e{\varepsilon}
\def\h{\eta}

\def\th{\theta}

\def\k{\kappa}
\def\m{\mu}

\def\r{\rho}

\def\nin{\noindent}

\def\Blb{\left\{}
\def\Brb{\right\}}

\def\giv{\,|\,}

\def\non{\nonumber}

\def\Eq{\ =\ }
\def\Def{\ :=\ }
\def\mm{{\mathcal M}}
\def\oar{{\emptyset \ne A \subsetneq [R]}}
\def\Le{\ \le\ }

\def\sjn{\sum_{j=1}^n}
\def\sjb{\sum_{j=1}^b}
\def\sji{\sum_{j\ge1}}

\def\Ref#1{{\rm (\ref{#1})}}

\def\bp{\begin{proof}}
\def\ep{\end{proof}}

\def\ignore#1{}
\def\ex{\E}

\def\HG{{\rm HG}\,}
\def\Po{{\rm Po}\,}

\def\Bi{{\rm Bi}\,}
\def\law{{\cal L}}

\def\srr{\sum_{r=1}^R}
\def\uk{(\th_k)}
\def\lbb{_{(b)}}
\def\MN{{\rm MN}\,}
\def\sscj{\sum_{s=1}^{C_j}}
\def\hK{{\widehat K}}
\def\Bl{\left(}
\def\Br{\right)}

\begin{document}
\alphafootnotes
\author[A. D. Barbour and S. Tavar\'e]
{Andrew D. Barbour\footnotemark\
and Simon Tavar\'e\footnotemark}
\chapter[Cancer genomes]{Assessing molecular variability in cancer genomes}
\footnotetext[1]{Institut f\"ur Mathematik, Universit\"at Z\"urich,
  Winterthurerstrasse 190, CH--8057 Z\"urich, Switzerland;
  A.D.Barbour@math.uzh.ch}
\footnotetext[2]{DAMTP and Department of Oncology, Cancer Research UK Cambridge
  Research Institute, Li Ka Shing Centre, Robinson Way, Cambridge CB2 0RE;
  st321@cam.ac.uk}
\arabicfootnotes
\contributor{Andrew D. Barbour
  \affiliation{University of Z\"urich}}
\contributor{Simon Tavar\'e
  \affiliation{University of Cambridge}}

\renewcommand\thesection{\arabic{section}}
\numberwithin{equation}{section}
\renewcommand\theequation{\thesection.\arabic{equation}}
\numberwithin{figure}{section}
\renewcommand\thefigure{\thesection.\arabic{figure}}
\numberwithin{table}{section}
\renewcommand\thetable{\thesection.\arabic{table}}

\begin{abstract}
The dynamics of tumour evolution are not well understood. In this paper we provide a statistical framework for evaluating the molecular variation observed in different parts of a colorectal tumour. A multi-sample version of the Ewens Sampling Formula forms the basis
for our modelling of the data, and we provide  a simulation
procedure for use in obtaining reference distributions for the statistics of interest.  We also describe the large-sample asymptotics of the joint distributions of the variation observed in different parts of the tumour.  While actual data should be evaluated with reference to the simulation procedure, the asymptotics serve to provide theoretical guidelines, for instance with reference to the choice of possible statistics. 
\end{abstract}

\subparagraph{AMS subject classification (MSC2010)}92D20; 92D15, 92C50, 60C05, 62E17

\section{Introduction}

Cancers\index{cancer|(} are thought to develop as clonal expansions from a
single transformed, ancestral cell\index{cell|(}. Large-scale sequencing
studies have shown that cancer genomes\index{genome} contain somatic
mutations\index{mutation|(} occurring in many genes\index{gene};
cf. \citet{Greenman06,Sjoblom06,Shah09}\index{Greenman, C.}\index{Sjoblom, T.@Sj\"oblom, T.}\index{Shah, S. P.}. Many of these mutations are thought to
be passenger mutations\index{mutation!passenger mutation|(} (those that are
not driving the behaviour of the tumour),\index{tumour|(} and some are
pathogenic driver mutations that influence the growth of the tumour. The
dynamics of tumour evolution are not well understood, in part because serial
observation of tumour growth in humans is not possible.

In an attempt to better understand tumour growth and structure, a number of
evolutionary approaches have been described.
\index{Merlo, L. M. F.}\index{Pepper, J. W.}\index{Reid, B. J.}\citet{Merlo06}
give an excellent overview of the field.
\citet{Tsao00}\index{Tsao, J. L.}\index{Yatabe, Y.|(} used non-coding
microsatellite loci as molecular tumour clocks in a number of human mutator
phenotype colorectal tumours. Stochastic models of tumour growth and
statistical inference\index{statistics} were used to estimate ancestral
features\index{ancestral history} of the tumours, such as their age (defined
as the time to loss of mismatch repair).
\citet{Campbell08}\index{Campbell, P. J.} used deep sequencing of a
\index{DNA (deoxyribonucleic acid)}DNA region to characterise the
phylogenetic relationships among clones within patients with B-cell chronic
lymphocytic leukaemia.
\citet{SiegmundPNAS}\index{Siegmund, K. D.|(}\index{Marjoram, P.|(}\index{Woo YJ@Woo, Y.-J.|(}\index{Shibata, D.|(}
used passenger mutations at particular CpG sites to infer aspects of the
evolution of colorectal tumours in a number of patients, by examining the
methylation\index{methylation|(} patterns in different parts of each tumour.

The problem of comparing the \Index{molecular variation} present in different
parts of a tumour is akin to the following problem from population
genetics\index{mathematical genetics}. Suppose that $R$
observers\index{multiple observers} take
samples\index{sampling|(} of sizes $n_1$, \ldots, $n_R$ from a population, and
record the molecular variation seen in each member of their sample. If the
population were indeed homogeneous, it makes sense to ask about the relative
amount of \Index{genetic variation} seen in each sample. For example, how many
genetic types are seen by all the observers, how many are seen by a single
observer, and so on.
\citet{wje07}\index{Ewens, W. J.}\index{RoyChoudhury, A.}\index{Lewontin, R. C.}\index{Wiuf, C.} discuss this problem in the case of $R = 2$
observers; the methodological contribution of the present paper addresses the
case of multiple observers. The theory is used to study the spatial
organization of the colorectal tumours studied in~\citet{SiegmundPNAS}.

This paper is organized as follows. In Section \ref{sec:cangen} we describe
the tumour
\index{data!colorectal cancer data|(}data that form the motivation for our work.
The Ewens Sampling
Formula\index{Ewens, W. J.!Ewens sampling formula (ESF)}, which forms the basis for
our modelling of the data, is described in Section \ref{sec:esf}, together
with a \Index{simulation} procedure for use in obtaining reference
distributions for the statistics of interest. The procedure for testing
whether the observers are homogeneous among themselves is illustrated in
Section \ref{analysis}.  The remainder of the paper is concerned with the
large-sample asymptotics of the joint distributions of
the \index{allele|(}allele counts from the different observers.  While actual
data should be evaluated with reference to the simulation procedure, the
asymptotics serve to provide theoretical guidelines, for instance with
reference to the choice of possible statistics.

\section{Colorectal cancer data}\label{sec:cangen}

In this section we describe the colorectal cancer data that motivate the
ensuing work. \citet{Yatabe01}\index{Yatabe, Y.|)}\index{Shibata, D.} describe
an experimental procedure for sampling\index{sampling|)} CpG DNA methylation
patterns from cells. These methylation patterns change during cell division,
due to random mutational events that result in switching an unmethylated site
to a methylated one, or vice versa. The methylation patterns obtained from a
particular locus may be represented as strings of binary outcomes, a 1
denoting a methylated site and a 0 an unmethylated one.

\begin{figure}
	\begin{center} \resizebox{0.49\textwidth}{!}{\includegraphics{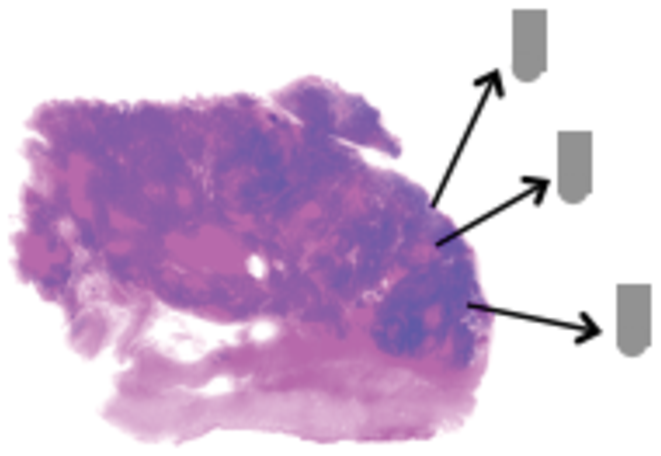}} \resizebox{0.49\textwidth}{!}{\includegraphics{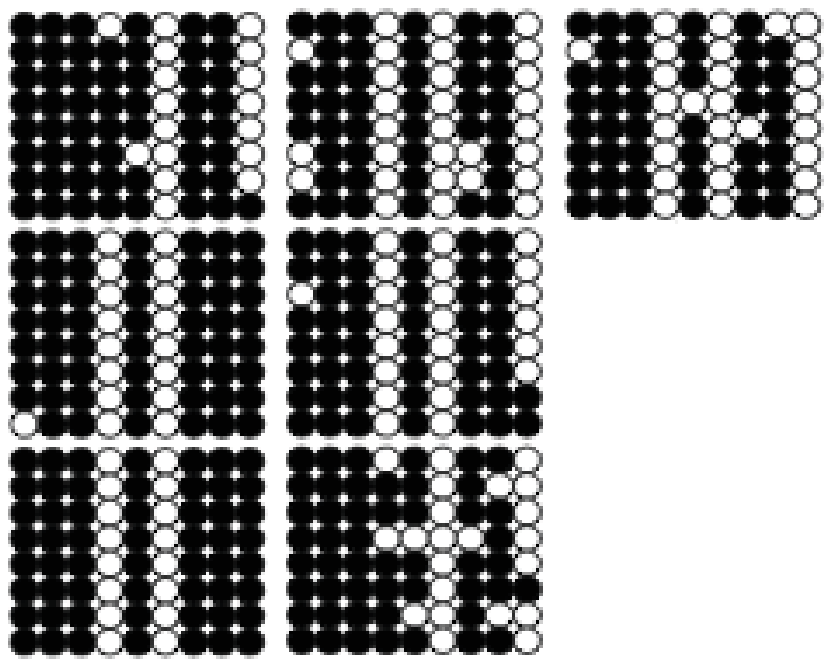}} \end{center} \caption[short
		title]{Left panel: sampling illustrated from three glands from
		one side of a colorectal tumour. Each gland contains
		2,000--10,000 cells. Right panel: Methylation data from the
		BGN locus from 7 glands from the left side of Cancer 1 (CNC1,
		from \cite{SiegmundPNAS}). 8 cells are sampled from each
		gland. Each row of 9 circles represents the methylation
		pattern in a cell. Solid circles denote methylated sites, open
		circles unmethylated. See Table \ref{tab1} for further
		details.\label{fig:can1}}
\end{figure}

\citet{SiegmundPNAS} studied 12 human colorectal tumours, each taken from
male patients of known ages. Samples of cells were taken from 7 different
glands from each of two sides of each tumour, and the methylation pattern at
two neutral (passenger) CpG loci (BGN, 9 sites; and LOC, 14 sites; both are on
the X \Index{chromosome}) was measured in each of 8 cells from each
gland. Figure \ref{fig:can1} illustrates the sampling, and depicts the data
from the left side of Cancer~1.

Data obtained from methylation patterns may be compared in several ways. We
focus on the simplest method that considers whether or not cells have
the same allele (that is, an identical pattern of 0s and 1s). Here we do not
exploit information about the detailed structure of the methylation patterns,
for which the reader is referred to \cite{SiegmundPNAS}. In Table \ref{tab1}
we present the data from Cancer 1 shown in Figure \ref{fig:can1} in a
different way. The body of the table shows the numbers of cells of each allele
(or type) in each of the 7 samples. The third row of the Table shows the
numbers $K_i$ of different alleles seen in each sample. In Table \ref{tab2} we
give a similar breakdown for data from the left side of Cancer 2.

The last column in Tables \ref{tab1} and \ref{tab2} gives the combined
distribution of allelic variation at this locus in the two
tumours. Qualitatively, the two tumours seem to have rather different
behaviour: Cancer 1 has far fewer alleles than Cancer 2, and their allocation
among the different samples is more homogeneous in Cancer 1 than in
Cancer 2. In the next sections we develop some theory that allows us to
analyse this variation more carefully.\index{mutation!passenger mutation|)}

\begin{table}
\begin{minipage}{100mm}
\caption{Data for Cancer 1. 13 alleles were observed in the 7 samples. Columns
labelled 1--7 give the distribution of the alleles observed in each sample, and
column 8 shows the combined data. Data from cancer CNC1 in\/
{\rm \cite{SiegmundPNAS}}.}\label{tab1}
\end{minipage}
\addtolength\tabcolsep{2pt}
\begin{tabular}{r|rrrrrrr|r}\hline
$i$ & 1 & 2 & 3 & 4 & 5 & 6 & 7 & 8\\
\hline 
$n_i$ &8  & 8&   8&  8&   8& 8& 8 & 56\\
$K_i$ & 4 & 2 & 1 & 3 & 3 & 5 & 5 & 13\\
$\hat\theta_i$ & 2.50 & 0.49 & 0.00 & 1.25 & 1.25 & 4.69 & 4.69 & 5.01\\
\hline 
allele &&&&&&&& \\
1 & 1 &    &    & 5 & 5 & 1  & 4 & 16 \\
2 & 5 &    &    &    &    &  3 &    &  8 \\
3 & 1 &    &    &    &    &     &    & 1\\
4 & 1 &    &    &    &    & 1  &    & 2\\
5 &    & 7 & 8 &    & 2 &     &    & 17 \\
6 &    &  1 &   &    &    &     &    &  1 \\
7 &    &     &   &  2  &    &     &    &  2 \\
8 &    &     &   &  1&   1&    &   1& 3 \\
9 &    &    &   &     &    &  2 &     & 2\\
10&   &    &   &     &    &  1 &     & 1\\
11&   &    &   &     &    &     &  1 & 1\\
12&   &    &   &     &    &     &  1 & 1\\
13&   &    &   &     &    &     &  1 & 1\\
\hline
\end{tabular}
\end{table}

\begin{table}
\begin{minipage}{100mm}
\caption{Data for Cancer 2.  27 alleles were observed in the 7
samples. Columns labelled 1--7 give the distribution of the alleles observed in
each sample, and column 8 shows the combined data. Data from cancer COC1 in\/
{\rm \cite{SiegmundPNAS}}.}\label{tab2}
\end{minipage}
\addtolength\tabcolsep{1.5pt}
\begin{tabular}{r|rrrrrrr|r}\hline
$i$ & 1 & 2 & 3 & 4 & 5 & 6 & 7 & 8\\
\hline 
$n_i$ &8  & 8&   8&  8&   8& 8& 8 & 56\\
$K_i$ & 7 & 7 & 4 & 3 & 7 & 6 & 4 & 27\\
$\hat\theta_i$ & 23.11 & 23.11 & 2.50 & 1.25 & 23.11 & 9.23 & 2.50 & 19.88\\
\hline 
allele &&&&&&&& \\
1 & 1 &    &    &   &   &   &  & 1 \\
2 & 1 &    &    &    &    &   &    &  1 \\
3 & 2 & 1   & 2   & 1   &    &     &    & 6 \\
4 & 1 &    &    &    &    &   &    & 1 \\
5 &  1  & 1 &  &    &  &     &    & 2 \\
6 &   1 &   &   &    &    &     &    &  1 \\
7 &    1&     &   &    &    &     &    &  1 \\
8 &    &  1   &  4 &  4&   &   3 &   & 12 \\
9 &    &   1 &   &     &    &   &     & 1 \\
10&   &   2 &   &     &    &   &     & 2\\
11&   &  1  &   &     &    &     &   & 1\\
12&   &  1  &   &     &    &     &   & 1\\
13&   &    &  1 &     &    &  1   &   & 2\\
14 &  &    &   1 &  &  &   &  & 1 \\
15&  &    &    &  3  &    &   &    &  3 \\
16 &  &    &    &    &   1 &     &    & 1\\
17 &  &    &    &    &   2 &   &  1  & 3\\
18 &    &   &  &    &  1&     &    & 1 \\
19 &    &   &   &    &   1 &     &  1  &  2 \\
20 &    &     &   &    &  1  &     &    &  1 \\
21 &    &     &   &    &   1 &    &    & 1 \\
22 &    &    &   &     &    1&   &     & 1\\
23&   &    &   &     &    &  1 &     & 1\\
24&   &    &   &     &    &   1  &   & 1\\
25&   &    &   &     &    &  1   &  4 & 5\\
26&   &    &   &     &    &  1   &   & 1\\
27&   &    &   &     &    &     &  2 & 2\\
\hline
\end{tabular}
\end{table}

\section{The Ewens sampling formula}\label{sec:esf}
\index{Ewens, W. J.!Ewens sampling formula (ESF)|(}
Our focus is on identifying whether the data are consistent with a uniformly
mixing collection of tumour cells that are in approximate stasis, or are more
typical of patterns of growth such as described in
\citet{Siegmund08,SiegmundPNAS,SiegmundCC}\index{Siegmund, K. D.|)}\index{Marjoram, P.|)}\index{Woo YJ@Woo, Y.-J.|)}\index{Shibata, D.|)}.
Whatever the model, the basic ingredients that must be specified include how
the cells are related, the details of which depend on the
\Index{demographic model} used to describe the tumour evolution, and the
mutation process that describes the methylation patterns. A review is provided
in \citet{SiegmundCC}. We use a simple null model in which the population of
cells is assumed to have evolved for some time with an approximately constant,
large size of $N$ cells, the constancy of cell numbers mimicking stasis in
tumour\index{tumour|)} growth. The mutation model assumes that in each cell
division there is probability $u$ of a mutation resulting in a type that has
not been seen before---admittedly a crude approximation to the nature of
methylation\index{methylation|)} mutations arising in our sample. The
mutations are assumed to be neutral, a reasonable assumption given that the
BGN gene is expressed in connective tissue but not in the epithelium.  Thus
our model is a classical one from population
genetics\index{mathematical genetics!neutral model}, the so-called
infinitely-many-neutral-alleles
model\index{mathematical genetics!infinitely-many-alleles model}.

Under this model the distribution of the types observed in the combined data (i.e., the allele counts derived from the right-most columns of data from Tables \ref{tab1} and \ref{tab2}) has a distribution that depends on the parameter $\theta = 2 N u$. This distribution is known as the Ewens Sampling Formula \cite{wje72}, denoted by ESF($\theta)$, and may be described as follows. For a sample of $n$ cells, we write $(C_1,C_2,\ldots,C_n)$ for the vector of counts given by
$$
C_j = \mbox{number of types represented } j \mbox{ times in the sample},
$$
where $C_1 + 2 C_2 + \cdots + n C_n = n$. For the Cancer 1 sample we have  $n = 56$ and
$$
C_1 = 6, C_2 = 3, C_3 = 1, C_8 = 1, C_{16} = 1, C_{17} = 1,
$$ 
whereas for Cancer 2 we also have $n = 56$, but 
$$
C_1 = 17, C_2 = 5, C_3 = 2, C_5 = 1, C_6 = 1, C_{12} = 1.
$$
The distribution ESF($\theta$) is given by
\begin{equation}\label{esf1}
\pr[C_1 = c_1,\ldots,C_n = c_n] = \frac{n!}{\theta_{(n)}}\,\prod_{j=1}^n \left(\frac{\theta}{j}\right)^{c_j}\,\frac{1}{c_j!},
\end{equation}
for $c_1 + 2 c_2 + \cdots + n c_n = n$ and
$\theta_{(n)}:= \th(\th+1)\ldots(\th+n-1)$. An explicit connection between
mutations\index{mutation|)} resulting in the ESF and the
\Index{ancestral history} of
the individuals (cells) in the sample is provided by Kingman's
coalescent \citep{jfck82a,jfck82c}, and the connection with the infinite
population limit is given in
\citet{jfck82b,jfck93}\index{Kingman, J. F. C.!Kingman coalescent}.

We recall from \citep{wje72}\index{Ewens, W. J.} that $K = K_n := C_1 + \cdots
+ C_n$, the number of types in the sample, is a sufficient
statistic\index{sufficient statistic|(} for $\theta$, and that the
maximum-likelihood
estimator\index{maximum likelihood!maximum-likelihood estimator (MLE)} of
$\theta$ is the solution of the equation
\begin{equation}\label{mle}
K_n = \ex_\theta(K_n) = \sum_{j=1}^n \frac{\theta}{\theta+j-1}.
\end{equation}
The conditional distribution of the counts $C_1$, \ldots, $C_n$ given $K_n$
does not depend on $\theta$, and thus may be used to assess the goodness-of-fit
of the model\index{goodness of fit}.

\subsection{The multi-observer ESF}\index{multiple observers|(}
So far, we have described the distribution of variation in the entire sample,
rather than in each of the subsamples from the different glands
separately. The joint law of the counts of different alleles seen in the $R$
glands (that is, by the $R$ observers) is precisely that obtained by taking a
hypergeometric sample\index{sampling!hypergeometric sampling} of sizes $n_1$,
$n_2$, \ldots, $n_R$ from the $n$ cells in the combined sample.  It is a
consequence of the consistency property of the ESF that the sample seen by
each observer $i$ has its own ESF, with parameters $n_i$ and $\theta$, $i=1$,
2, \ldots, $R$. Tables \ref{tab1} and \ref{tab2} give the observed values for
the two tumour\index{tumour|(} examples.

We are interested in assessing the goodness-of-fit of the tumour data
subsamples to our simple model of a homogeneous tumour in stasis. Because
$K_n$ is sufficient\index{sufficient statistic|)} for $\theta$ in the combined
sample, this can be performed by using the joint distribution of the counts
seen by each observer, conditional on the value of $K_n$. To simulate from
this distribution we use the Chinese Restaurant Process, as described in the
next section.

\subsection{The Chinese Restaurant Process}\index{Chinese restaurant process (CRP)|(}
We use simulation to find the distribution of certain test statistics relating
to the multiple observer data. To do this we exploit a simple way to simulate
a sample of individuals (cells\index{cell|)} in our example) whose allele
counts follow the ESF($\theta$). The method, known as the Chinese Restaurant
Process (CRP), after \citet{dp86}\index{Diaconis, P.}\index{Pitman, J. [Pitman, J. W.]},
simulates individuals in a sample sequentially. The first individual is given
type 1. The second individual is either a new type (labelled 2) with
probability $\theta/(\theta+1)$, or a copy of the type of individual 1, with
probability $1/(\theta+1)$. Suppose that $k-1$ individuals have been assigned
types. Individual $k$ is assigned a new type (the lowest unused positive
integer) with probability $\theta/(\theta+k-1)$, or is assigned the type of
one of individuals 1, 2, \ldots, $k-1$ selected uniformly at
random. Continuing until $k = n$ produces a sample of size $n$, and the joint
distribution of the number of types represented once, twice, \ldots\ is indeed
ESF($\theta$).

Once the sample of $n$ individuals is generated, it is straightforward to
subsample without replacement to obtain $R$ samples, of sizes $n_1$, \ldots,
$n_R$, in each of which the distribution of the allele counts follows the
ESF($\theta$) of the appropriate size. This may be done
sequentially\index{sampling!sequential sampling}, choosing $n_1$ without replacement to
be the first sample, then $n_2$ from the remaining $n - n_1$ to form the
second sample, and so on.

When samples of size $n$ are required to have a given number of alleles, say
$K_n = k$, this is most easily arranged by the \Index{rejection method}: the
CRP is run to produce an $n$-sample, and that run is rejected unless the
correct value of $k$ is observed. Since conditional on $K_n = k$ the
distribution of the allele frequencies is independent of $\theta$, we have
freedom to choose $\theta$, which may be taken as the MLE $\hat\theta$
determined in (\ref{mle}) to make the rejection probability as small as
possible.

\section{Analysis of the cancer data}\label{analysis}

We have noted that the combined data in the $R$-observer ESF have the
ESF($\theta$) distribution with sample size $n = n_1+\cdots+n_R$, while the
$i$th observer's sample has ESF($\theta$) distribution with sample size
$n_i$. Of course, these distributions are not independent.  To test whether
the combined data are consistent with the ESF, we may use a statistic
suggested by \citet{gaw78b}\index{Watterson, G. A.}, based on the distribution
of the sample \Index{homozygosity}
$$
F = \sum_{j=1}^n C_j\,\left( \frac{j}{n}\right)^2
$$
found after conditioning on the number of types seen in the combined sample. Each marginal sample may be tested in a similar way using the appropriate value of $n$.

Since our cancer data arise as the result of a spatial
sampling\index{sampling!spatial sampling} scheme, it is natural to consider statistics
that are aimed at testing whether the samples can be assumed homogeneous, that
is, are described by the multi-observer ESF.  Knowing the answer to this
question would aid in understanding the dynamics of tumour evolution, which in
turn has implications for understanding \Index{metastasis} and response to
therapy.

To assess this, we use as a simple illustration  the sample variance of the numbers of types seen in each sample. The statistic may be written as
\begin{equation}\label{kvar}
   Q  \Def \frac{1}{R-1}\,\sum_{i=1}^R (K_i - \bar K)^2 
     \Eq \frac{1}{R(R-1)} \sum_{1 \leq i < j \leq R} (K_i - K_j)^2,
\end{equation}
the latter expression emphasizing its role as a measure of the average discrepancy
between samples. In the next paragraphs, we discuss the structure of Cancers 1 and 2 using these statistics.

\paragraph{Cancer 1}
We begin with a comparison of the data from the two sides of Cancer 1. In this
case $n_1 = 56, n_2 = 56$ and the combined sample of $n = 112$ has $K_{112} =
16$ and $F = 0.237$. The 5th and 95th percentiles of the null distribution of
$F$ found by the conditional CRP\index{Chinese restaurant process (CRP)|)}
simulation described in the last section are 0.108 and 0.277 respectively,
suggesting no anomaly with the underlying ESF model. For the left side of the
cancer (Table \ref{tab1}), $K_{56} = 13$ and $F = 0.209$, while for the right
side (data not shown), $K_{56} = 10$ and $F = 0.293$.  In both cases these
observed values of $F$ are consistent with the ESF.  We then use the statistic
$Q$ to investigate whether the data from the 7 glands from the left side of
the tumour are homogeneous. We observed $Q = 2.24$, and the null distribution
of $Q$ can also be found from the conditional CRP \Index{simulation}.  We
obtained 5th and 95th percentiles of 0.29 and 2.48 respectively, supporting
the conclusion of a homogeneous tumour.

\paragraph{Cancer 2} The comparison of the two sides of Cancer 2  is more interesting. Once more $n_1 = 56, n_2 = 56$ but the combined sample of $n = 112$ now has $K_{112} = 48$ and  $F = 0.081$. The 99th percentile of the null distribution of $F$  is 0.060, suggesting that the ESF model is not adequate to describe the combined data. At first glance the anomaly can be attributed to the data from the right side of the tumour (not shown here), for which $K_{56} = 29$ and  $F =  0.105$, far exceeding the 99th percentile of 0.089. For the left side (Table \ref{tab2}), $F = 0.083$, just below the 95th percentile of 0.084. Thus the left side seems in aggregate to be adequately described by the ESF model. Further examination of the data from the 7 glands reveals a different story.
From the third row of Table \ref{tab2} we calculate $Q  = 2.95$, far exceeding
the estimated 99th percentile of 2.33. Thus a more detailed view of the way
the mutations\index{mutation|(} are shared among the glands shows that these data are indeed
inconsistent with the homogeneity expected in the multi-observer ESF.
\index{Ewens, W. J.!Ewens sampling formula (ESF)|)}



\bigskip
Of course, many other statistics could have been considered.  A natural starting point for constructing them would be the numbers of alleles that are seen only by a specific subset $A$ of the observers, where $A$ ranges over the $2^R-2$ non-empty proper subsets of the $R$ observers. Such statistics form the basis of the results in Section~\ref{sec:1.5}.

Rejection of the null hypothesis of the uniformly mixing homogeneous tumour
model can occur for many reasons, for example because of non-uniform mutation
rates\index{mutation!rate}, different \Index{demography} of \Index{cell}
growth, non-neutrality\index{mathematical genetics!neutral model} of the
mutations (which might apply to the BGN locus if in fact it were expressed in
tissue in the tumour\index{tumour|)}), and unforeseen effects of the simple
mutation\index{mutation|)} model itself.  Which of these hypotheses is most
likely requires a far more detailed analysis of competing models, as for
example outlined in
\cite{Siegmund08,SiegmundPNAS,SiegmundCC}\index{Siegmund, K. D.}\index{Marjoram, P.}\index{Woo YJ@Woo, Y.-J.}\index{Shibata, D.}.\index{data!colorectal cancer data|)}

\section{Poisson approximation}\label{sec:1.5}\index{Poisson, S. D.!Poisson approximation|(}

In this section, we derive Poisson approximations to the joint distribution of
the numbers of alleles that are seen only by specific subsets $A$ of the
observers.  As mentioned above, functionals of these counts can be used as
statistics to test for the homogeneity of (subgroups of) observers. Our
approximations come together with bounds on the total variation
distance\index{total-variation metric|(}
between the actual and approximate distributions.  We begin with the case of
$R = 2$ observers, and with the statistic $K_1-K_2$.

\subsection{2 observers}
We write $C := (C_1,C_2,\ldots)$, where $C_j = 0$ for $j > n$, and recall 
Watterson's\index{Watterson, G. A.} result, that  $(C_1,\ldots,C_{n})$ are jointly distributed
according to $\law(Z_1,Z_2,\ldots,Z_{n} \giv T_{0n}(Z) = n)$, where
$(Z_j,\,j\ge1)$ are independent with $Z_i \sim \Po(\th/i)$, and
\eq\label{T-not}
   T_{rs}(c) \Eq \sum_{j=r+1}^s jc_j, \qquad c \in \integ_+^\infty,
\en
\citep{gaw74a}.
The sampled individuals can be labelled 1 or 2, according to which observer
sampled them; under the above model, the $n_1$ 1-labels and $n_2$ 2-labels
are distributed at random
among the individuals, irrespective of their allelic type.  Let $K_r$ denote
the number of distinct alleles observed by the $r$-th observer, $r=1$, 2.
\citet{wje07}\index{Ewens, W. J.}\index{RoyChoudhury, A.}\index{Lewontin, R. C.}\index{Wiuf, C.} observed that, in the case $n_1=n_2$ and for
large $n$, $(K_1-K_2)/\log n$ is equivalent to
the difference in the estimates of the mutation\index{mutation!rate} rate made by the two
observers. The same is asymptotically true also as $n$ becomes large,
if $n_1/n \sim p_1$ for any fixed $p_1$.  This motivates us to look for
a distributional approximation to the distribution of the difference
$K_1-K_2$.

\begin{theorem}\label{k1-k2}
For any $n_1$, $n_2$ and~$b$,
\[
  \dtv(\law(K_1-K_2),\law(P_1-P_2)) 
     \Le \frac{kb}{n-1} + \frac{k'\r^{b+1}}{(b+1)(1-\r)},
\]
for suitable constants $k$ and $k'$,
where $P_1$ and $P_2$ are independent Poisson random variables having
means $\th\log\{1/(1-p_1)\}$ and $\th\log\{1/(1-p_2)\}$ respectively,
with $p_r := n_r/n$,
and where $\r = \max\{1-p_1,1-p_2\}$. The choice $b=b_n= \lfloor
\log n/\log(1/\rho) \rfloor$ gives a bound
of order $O\bigl(\log n / (n\min\{p_1,p_2\})\bigr)$.
\end{theorem}     

\bp
Group the individuals in the combined sample according to their
allelic type, and let $M_{js}$ denote the number of individuals
that were observed by observer 1 in the 
$s$-th of the $C_j$ groups of size $j$,
the remaining $j-M_{js}$ being observed by observer 2.  Define
\[
   S_j^1\ :=\ \sum_{s=1}^{C_j} I[M_{js}=j]\quad\mbox{and}\quad 
   S_j^2\ :=\ \sum_{s=1}^{C_j} I[M_{js}=0]
\]
to be the numbers of $j$-groups observed only by observers 1 and 2,
respectively.  Then it follows that 
\[
   K_1 - K_2 \Eq S^1-S^2,
\] 
where $S^r := \sum_{j=1}^{n} S_j^r$ .
The first step in the proof is to show that the effect of the large
groups is relatively small.
  
Note that the probability that an allele which is
present $j$ times in the combined sample was \emph{not} observed by
observer 1 is
\[
   \prod_{i=0}^{j-1} \frac{n_1-i}{n-i} \Le (1-p_1)^j;
\]
similarly, the probability that it was not observed by observer 2 is at most
$(1-p_2)^j$.   Hence, conditional on $C$, the probability that  any of the alleles
present more than $b$ times in the combined sample is seen by \emph{only one}
of the observers is at most
\begin{align*}
  \ex\Biggl\{\, \sum_{j=b+1}^{n}(S_j^1 + S_j^2)\,\Bigg|\, C \Biggr\} &\Le
   \sum_{j=b+1}^{n} C_j\{(1-p_1)^j + (1-p_2)^j\}\\ &\Le 2\sum_{j=b+1}^{n} \r^jC_j,
\end{align*}
whatever the value of $b$.  Hence, writing $U_b := \sum_{j=1}^b (S_j^1-S_j^2)$,
we find that
\eq
   \pr[K_1-K_2 \neq U_b] \Le 2\sum_{j=b+1}^{n} \r^j\ex C_j 
   \Le \frac{2k_1\r^{b+1}}{(b+1)(1-\r)},  \label{small-is-OK}
\en
where, by Watterson's  formula \cite{gaw74a}\index{Watterson, G. A.|(} for the means of the component sizes,
we can take $k_1 := (2\th+e^{-1})$ if $n\ge 4(b+1)$ (and $k_1:=\th$ if $\th\ge1$). 

To approximate the distribution of $U_b$, note that, conditional on $C$, the
number of 1-labels among the individuals in allele groups of at most $b$
individuals has a \Undex{hypergeometric distribution}
\[
    \HG(T_{0b}(C);n_1;n),
\]
where $\HG(s;m;n)$ denotes the number of black balls obtained in $s$ draws from
an \Index{urn} containing $m$ black balls out of a total of $n$.  
By Theorem 3.1 of \citet{dh04}\index{Holmes, S.}, we have
\eq\label{holmes}
  \dtv(\HG(T_{0b}(C);n_1;n),\Bi(T_{0b}(C),p_1)) \Le \frac{T_{0b}(C)-1}{n-1}.
\en 
Hence, conditional on $C$, the joint distribution of labels among individuals
differs in total variation from that obtained by independent Bernoulli\undex{Bernoulli, J.} random 
assignments, with label 1 having probability $p_1$ and label 2 probability $p_2$,
by at most $(T_{0b}(C)-1)/(n-1)$.

Now, by Lemma 5.3 of \citet{ABT03}\index{Arratia, R.}, we also have
\[
  \dtv(\law(C_1,\ldots,C_b),\law(Z_1,\ldots,Z_b)) \Le \frac{c_\th b}{n},
\]
with $c_\th \le 4\th(\th+1)/3$ if $n\ge4b$.  Hence, and from \Ref{holmes}, 
it follows that
\eqa
  &&\dtv(\law(C_1,\ldots,C_b; \{M_{js},\, 1\le s\le C_j,\,1\le j\le b\}),\non\\
   &&\qquad \qquad\qquad   
        \law(Z_1,\ldots,Z_b; \{N_{js},\, 1\le s\le Z_j,\,1\le j\le b\})) \non\\
   &&  \qquad\Le  \frac{c_\th b}{n} +  \frac{\ex(T_{0b}(C))-1}{n-1}, 
    \label{detailed}
\ena       
where $(N_{js};\,s\ge1,\,1\le j\le b)$ are independent of each other and of
$Z_1$, \ldots, $Z_b$, with $N_{js} \sim \Bi(j,p)$.  But now the values
of the $N_{js}$, $1\le s\le Z_j$, $1\le j\le b$, can be interpreted as the 
numbers of 1-labels assigned to each of $Z_j$ groups of size $j$ for each
$1\le j\le b$, again under independent Bernoulli\undex{Bernoulli, J.} random 
assignments, with label 1 having probability $p_1$ and label 2 probability $p_2$.
Hence, since the $Z_j$ are independent Poisson random variables, the counts 
\[
   T_j^1\ :=\ \sum_{s=1}^{Z_j} I[N_{js}=j]\quad\mbox{and}\quad 
   T_j^2\ :=\ \sum_{s=1}^{Z_j} I[N_{js}=0]
\] 
are pairs of independent Poisson distributed random variables, with
means $\th j^{-1} p_1^j$ and $\th j^{-1}p_2^j$, and are also independent
of one another.  Hence it follows that
\eq\label{partial-poisson}
   V_b\ :=\ \sum_{j=1}^b (T_j^1-T_j^2) \ \sim\ P_{1b} - P_{2b},
\en
where $P_{1b}$ and $P_{2b}$ are independent Poisson random variables,
with means $\th\sum_{j=1}^b j^{-1}p_1^j$ and $\th\sum_{j=1}^b j^{-1}p_2^j$, 
respectively.  Comparing the def\-initions of $U_b$ and $V_b$, and 
combining \Ref{detailed} and \Ref{partial-poisson}, it thus
follows that
\eq\label{U-approx}
   \dtv(\law(U_b),\law(P_{1b} - P_{2b})) \Le \frac{(c_\th+k_2)b}{n-1},
\en
with $k_2 = 4\th/3$ for $n\ge4b$, once again by Watterson's formula \cite{gaw74a}\index{Watterson, G. A.|)}.

With \Ref{small-is-OK} and \Ref{U-approx}, the argument is all but complete;
it simply suffices to observe that, much as in proving \Ref{small-is-OK},
\[
   \dtv(\law(P_1),\law(P_{1b})) + \dtv(\law(P_2),\law(P_{2b})) \Le  
    \frac{2\th \r^{b+1}}{(b+1)(1-\r)};
\] 
we take $k := 4\vee(c_\th+k_2)$ and $k' := 2(\th + k_1)$.
\ep

\subsection{$R$ observers}
The proof of Theorem \ref{k1-k2} actually shows that the joint distribution
of $S^1$ and $S^2$, the numbers of types seen respectively by observers 1 
and 2 alone, is close to that of independent Poisson random variables $P_1$
and $P_2$.
For $R\ge3$ observers, we use a similar approach 
to derive an approximation to the joint distribution of the numbers of alleles\index{allele|)}
seen by each proper subset $A$ of the $R$ observers. 

Suppose that
the $r$-th observer samples $n_r$ individuals, $1\le r\le R$, and set
$n := \srr n_r$, $p_r := n_r/n$.  Define the component frequencies in
the combined sample as before, and set $M_{js} = m := (m_1,\ldots,m_R)$
if the $r$-th observer\index{multiple observers|)} sees $m_r$ of the $j$ individuals in the $s$-th
of the $C_j$ groups of size $j$.  For any $\oar$, where $[R] :=
\{1,2,\ldots,R\}$, define
\begin{align*}
   &\mm_{Aj} \Def\\ &{}\quad\Bigl\{m\in\integ_+^R\colon\, \srr m_r = j,\,
    \{r\colon\,m_r\ge1\} = A,\, \{r\colon\,m_r=0\} = [R]\setminus A\Bigr\},
\end{align*}
and set 
\[
      S_j^A \Def \sscj I[M_{js}\in A].
\]
Our interest lies now in approximating the joint distribution of the counts
$(S^A,\allowbreak \oar)$, where $S^A := \sjn S_j^A$.  To do so, we need a set
of independent Poisson random variables $(P^A,\,\oar)$, with
$P^A \sim \Po(\l^A(\th))$, where 
\eq\label{lambda-A-def}
   \l_j^A(\th) \Def \frac{\th}{j}\, \MN(j;p_1,\ldots,p_R)\{\mm_{Aj}\}
   \quad\mbox{and}\quad \l^A(\th) \Def \sji \l_j^A(\th);
\en
here, $\MN(j;p_1,\ldots,p_R)$ denotes the multinomial distribution 
with $j$ trials and cell probabilities $p_1$, \ldots, $p_R$.

\begin{theorem}\label{R-obs}
In the above setting, we have
\begin{align*}
   &\dtv\Bigl(\law((S^A,\,\oar)),\bigtimes_{\oar} \Po(\l^A(\th)) \Bigr)\\
    &{}\qquad\Le \frac{k_Rb}n + \frac{k_R'\r^{b+1}}{(b+1)(1-\r)},
\end{align*}
where $\r := \max_{1\le r\le R}(1-p_r)$. Again $b = b_n =  \lfloor
\log n/\log(1/\rho) \rfloor$ is a good choice.
\end{theorem}

\bp
The proof runs much as before.  First, the bound
\[
   \ex\Biggl\{ \sum_{j=b+1}^{n} \sum_{\oar} S_j^A\,\Bigg|\, C\Biggr\}
    \Le \sum_{j=b+1}^{n} C_j \srr (1-p_r)^j \Le R\sum_{j=b+1}^{n} \r^j C_j
\]
shows that 
\eq\label{cutoff-mv}
     \pr\left[ \bigcup_{\oar} \{S^A \ne S^A\lbb\} \right]
       \Le \frac{Rk_1\r^{b+1}}{(b+1)(1-\r)},
\en
where $S^A\lbb := \sjb S_j^A$.  Then, by Theorem 4
of \citet{df80}\index{Diaconis, P.}\index{Freedman, D. A.}, 
\[
   \dtv\Bigl(\HG(T_{0b}(C);n_1,\ldots,n_R;n),\MN(T_{0b}(C);p_1,\ldots,p_R)\Bigr)
      \Le \frac{RT_{0b}}n,
\]
from which it follows that
\eqa
  &&\dtv\Bigl(\law(C_1,\ldots,C_b; \{M_{js},\, 1\le s\le C_j,\,1\le j\le b\}),\non\\
   &&\qquad \qquad\qquad   
        \law(Z_1,\ldots,Z_b; \{N_{js},\, 1\le s\le Z_j,\,1\le j\le b\})\Bigr) \non\\
   &&  \qquad\Le  \frac{c_\th b}{n} +  \frac{R\ex(T_{0b}(C))}{n}, 
    \label{detailed-mv}
\ena       
where $(N_{js};\,s\ge1,\,1\le j\le b)$ are independent of each other and of
$Z_1$, \ldots, $Z_b$, with $N_{js} \sim \MN(j;p_1,\ldots,p_R)$. Then the random
variables
\[
     T_j^A \Def \sum_{s=1}^{Z_j} I[N_{js}\in A],\qquad\oar,
\]
are independent and Poisson distributed, with means $\l^A\lbb(\th) :=\break \sjb \l_j^A(\th)$,
and
\[
   \dtv\{\law(S^A\lbb,\,\oar),\,\law(T^A\lbb,\,\oar)\} \Le \frac{k_2'b}n,
\]
with $k_2' := c_\th + 4R\th/3$, where 
\[
     T^A\lbb \Def \sjb T^A_j \ \sim\ \Po(\l^A\lbb(\th)).
\]
Finally, much as before,
\[
     \dtv\Bigl(\law((T^A\lbb,\,\oar)),\,\bigtimes_{\oar} \Po(\l^A(\th)) \Bigr)
      \Le \frac{R\th\r^{b+1}}{(b+1)(1-\r)},
\]
and we can take $k_R := 4\vee(c_\th+Rk_2)$ and $k_R' := R(\th+k_1)$ in
the theorem.
\ep

We note that the Poisson means $\lambda_j^A(\theta)$ appearing in (\ref{lambda-A-def})
may be calculated using an \Index{inclusion-exclusion} argument. For reasons of
\Index{symmetry} it is only necessary to compute $\lambda_j^A(\theta)$ for sets of the
form $A = [r] = \{1,2,\ldots,r\}$ for $r = 1$, 2, \ldots, $R-1$. We obtain
\begin{equation}\label{eq5}
 {\rm MN}(j;p_1,\ldots,p_R)\{\mm_{[r]j} \} =
\sum_{l=1}^{r} (-1)^{r-l} \sum_{J \subseteq [r], |J|=l} \left(\sum_{u \in J} p_u\right)^j, 
\end{equation}
from which the terms $\lambda^A(\theta)$ readily follow as
\begin{equation}\label{eq6}
\lambda^{[r]}(\theta) =
- \theta\,\sum_{l=1}^{r} (-1)^{r-l} \sum_{J \subseteq [r], |J|=l} \log\left(1 - \sum_{u \in J} p_u\right). 
\end{equation}

\subsection{The conditional distribution}
In statistical applications, such as that discussed above, the value
of $\th$ is unknown, and has to be estimated. Defining
\[
   K_{st}(c) \Def \sum_{j=s+1}^t c_j,
\]
the quantity $K_{0n}(C)$ is sufficient\index{sufficient statistic} for $\th$,
and the null distribution appropriate for testing model
fit\index{goodness of fit} is then the conditional distribution
$$
    \law((S^A,\,\oar)\giv K_{0n}(C)= k), 
$$
where $k$ is the observed value of $K_{0n}(C)$.  Hence we need to approximate
this distribution as well.  Because of sufficiency, the distribution no longer
involves $\th$.  However, for our approximation, we shall need to define means
for the approximating Poisson random variables $P^A\sim \Po(\l^A(\th))$, as
in \Ref{lambda-A-def}, and these need a value of $\th$ for their
definition. We thus take $\l^A(\th_k)$ for our approximation, for convenience
with $\th_k := k/\log n$; the
MLE\index{maximum likelihood!maximum-likelihood estimator (MLE)}
given in \Ref{mle} could equally well have been used.

The proof again runs along the same lines.
Supposing that the probabilities $p_1$, \ldots, $p_R$ are
bounded away from $0$, we can take $b := b_n := \lfloor\log n / \log(1/\r)\rfloor$ 
in Theorem \ref{R-obs}, and use \Ref{cutoff-mv} to show that it is enough
to approximate $\law((S^A\lbb,\,\oar)\giv K_{0n}(C)= k)$.  Then, since
the arguments conditional on the whole realization $C$ remain the same when
restricting $C$ to the set $\{K_{0n}(C)=k\}$, it is enough to show that
the distributions 
$$
    \law(C_{[0,b]} \giv K_{0n}(C)=k)\quad \mbox{and}\quad \law_{\th_k}(Z_{[0,b]})
$$ 
are close enough, where
$c_{[0,b]} := (c_1,\ldots,c_b)$, to conclude that the Poisson approximation
of Theorem \ref{R-obs} with $\th=\th_k$ also holds conditionally
on  $\{K_{0n}(C)=k\}$.
Note also that the event $\{K_{0n}(C)=k\}$ has probability
at least as big as $c_1(\th_k)k^{-1/2}$ for some positive
function $c_1(\cdot)$, by  (8.17) of \citet{ABT03}\index{Arratia, R.|(}.

Defining $\l_{st}\uk := \sum_{j=s+1}^t j^{-1}\th_k$, we can now prove
the key lemma.

\begin{lemma}\label{conditioning-lemma}
   Fix any $\e,\h > 0$. Suppose that $n$ is large enough, so that $b+b^3 < n/2$.
Then there is a constant $\k$ such that, uniformly for 
$\e \le k/\log n \le 1/\e$, and for $c\in\integ_+^\infty$ with $K_{0b}(c) \le \h\log\log n$
and $T_{0b}(c) \le  b^{7/2}$,
\[
     \left| \frac{\pr[C_{[0,b]} = c_{[0,b]} \giv K_{0n}(C)=k]}
          {\pr_{\th_k}[Z_{[0,b]}=C_{[0,b]}]}  - 1 \right| \Le \frac{\k\log\log n}{\log n}.
\]
\end{lemma}

\bp
Since $\law(C_1,\ldots,C_n) = \law(Z_1,\ldots,Z_n \giv T_{0n}(Z)=n)$, it follows
that
\eqs
  \lefteqn{\pr[C_{[0,b]} = c_{[0,b]} \giv K_{0n}(C)=k]}\\
   &=& \frac{\pr[K_{0n}(C)=k \giv C_{[0,b]} = c_{[0,b]}]\pr[C_{[0,b]} = c_{[0,b]}]}
        {\pr[K_{0n}(C)=k]} \\
  &=& \frac{\pr[K_{bn}(C)=k - K_{0b}(c) \giv T_{0b}(C) = T_{0b}(c)]\pr[C_{[0,b]} = c_{[0,b]}]}
        {\pr[K_{0n}(C)=k]}.
\ens
We now use results from \S 13.10 of \citet{ABT03}.  First, as on p.~323,
\eqs
   \lefteqn{\pr_{\th_k}[K_{bn}(C)=k - K_{0b}(c) \giv T_{0b}(C) = T_{0b}(c)]} \\
     &=& \pr_{\th_k}[K_{bn}(Z)=k - K_{0b}(c) \giv T_{bn}(Z) = n - T_{0b}(c)],
\ens
and the estimate on p.~327 then gives
\eqa
   \lefteqn{\pr_{\th_k}[K_{bn}(Z)=k - K_{0b}(c) \giv T_{bn}(Z) = n - T_{0b}(c)]}\non\\
     &=& \Po(\l_{bn}\uk)\{k-K_{0b}(c)-1\}\,\{1 + O((\log n)^{-1}\log\log n)\},\phantom{X}
     \label{Poisson-1}
\ena
uniformly in the chosen ranges of $k$, $T_{0b}(c)$ and $K_{0b}(c)$, because of the choice
$\th=\th_k$.  Then
\eq\label{Poisson-2}
   \pr_{\th_k}[K_{0n}(C)=k] \Eq \Po(\l_{0n}\uk)\{k-1\}\,\{1 + O((\log n)^{-1})\},
\en
again uniformly in $k$, $T_{0b}(c)$ and $K_{0b}(c)$, by Theorem 5.4 of \citet{abt00}.  
Finally, 
\begin{align*}
     \left|\frac{\pr_{\th_k}[C_{[0,b]} = c_{[0,b]}]}
                {\pr_{\th_k}[Z_{[0,b]} = c_{[0,b]}]} - 1\right|
     &\Eq \left|\frac{\pr_{\th_k}[T_{bn}(Z) = n - T_{0b}(c)]}
                {\pr_{\th_k}[T_{bn}(Z) = n]} - 1\right|\\
     &\Eq O(n^{-1}b^{7/2}),
\end{align*}
by (4.43), (4.45) and Example 9.4 of \cite{ABT03}, if $b + b^3 < n/2$. The lemma now 
follows by considering the ratio of the
Poisson probabilities in \Ref{Poisson-1} and \Ref{Poisson-2}; note that
$\l_{0n}\uk - \l_{bn}\uk = O(\log\log n)$.
\ep

In order to deduce the main theorem of this section, we just need to bound
the conditional probabilities of the events $\{K_{0b}(C) > \h\log\log n\}$
and $\{T_{0b}(C) >  b^{7/2}\}$, given $K_{0n}(C) = k$.  For the first, note that
\eq\label{uncondition}
   \pr_{\th_k}[K_{0b}(C) > \h\log\log n] \Le
      \frac{c_{\th_k}b}n + \pr_{\th_k}[K_{0b}(Z) > \h\log\log n],
\en
and that $K_{0b}(Z) \sim \Po(\th_k\sjb j^{-1})$ with mean of order $O(\log\log n)$.
Hence there is an $\h$ large enough that
\[
   \pr_{\th_k}[K_{0b}(C) > \h\log\log n] \Eq O((\log n)^{-5/2}),
\]
uniformly in the given range of $k$.  Since also, from \Ref{Poisson-2},
\[
   \pr_{\th_k}[K_{0n}(C)=k] \ \ge\ \h'/\sqrt{\log n}
\]
for some $\h' > 0$, it follows immediately that
\eq\label{Kob-bnd}
   \pr_{\th_k}[K_{0b}(C) > \h\log\log n\giv K_{0n}(C)=k] \Eq
       O((\log n)^{-2}).
\en
The second inequality is similar.   We use the argument of \Ref{uncondition}
to reduce consideration to $\pr_{\th_k}[T_{0b}(Z) >  b^{7/2}]$, and  (4.44) of \citet{ABT03}
shows that
\[
   \pr_{\th_k}[T_{0b}(Z) > b^{7/2}]  \Eq O(b^{-5/2}) \Eq O((\log n)^{-5/2});
\]
the conclusion is now as for \Ref{Kob-bnd}.

In view of these considerations, we have established the following theorem,
justifying the Poisson approximation to the conditional distribution of the
$(S^A,\,\oar)$, using the estimated value $\th_k$ of $\th$ as parameter.

\begin{theorem}\label{conditional-satz}
   For any $0 <\e < 1$, uniformly in $\e \le k/\log n \le 1/\e$, we have
\begin{align*}
   &\dtv\Bigl(\law((S^A,\,\oar)\giv K_{0n}(C)=k),\bigtimes_{\oar} \Po(\l^A\uk)\Bigr)\\
     &{}\qquad\Eq O\Bigl(\frac{\log\log n}{\log n}\Bigr).
\end{align*}
\end{theorem}

\nin Note that the error bound is much larger for this approximation than
those in the previous theorems.  However, it is not unreasonable.  
From \Ref{cutoff-mv},
the joint distribution of the $S^A$ is almost entirely determined by that of
$C_1$, \ldots, $C_b$.  Now $\law(K_{0b}(C) \giv K_{0n}(C)=k)$ can be expected to
be close to $\law(K_{0b}(Z) \giv K_{0n}(Z)=k)$, which is binomial
$\Bi(k,p_{b,n})$, where
\[
   p_{b,n} \Def \frac{\sjb 1/j}{\sjn 1/j} \ \approx\ \frac{\log b}{\log n}
   \ \approx\ \frac{\log\log n}{\log(1/\r)\log n}.
\]
On the other hand, from Lemma 5.3 of \cite{ABT03}\index{Arratia, R.|)}, the {\it unconditional\/}
distribution of $K_{0b}(C)$ is very close to that of $K_{0b}(Z)$, a Poisson distribution.  The
total variation distance\index{total-variation metric|)} between the distributions $\Po(kp)$ and $\Bi(k,p)$
is of exact order $p$ if $kp$ is large (Theorem 2 of
\citet{bh84})\index{Hall, P. G.}.
Since $p_{b,n} \asymp \log\log n/\log n$, an error of this order in
Theorem \ref{conditional-satz} is thus in no way surprising.

\bigskip
We can now compute the mean $\m$ of the approximation to the distribution of
$Q$, as used in Section \ref{analysis}, obtained by using Theorem \ref{conditional-satz}.
We begin by noting that, using the theorem,
$$
   K_r-K_s \Eq \sum_{A\colon\,r\in A,s \notin A} S^A -
     \sum_{A\colon\,r\notin A,s \in A} S^A
$$
is close in distribution to
$$
  \hK_{rs} - \hK_{sr} \Def  
      \sum_{A\colon\,r\in A,s \notin A} P^A - \sum_{A\colon\,r\notin A,s \in A} P^A,
$$
where $P^A \sim \Po(\l^A\uk)$, $\oar$, are independent.  To compute the means
\[
    \l_{rs} \Def \sum_{A\colon\,r\in A,s \notin A} \l^A\uk
    \quad\mbox{and}\quad \l_{sr} \Def \sum_{A\colon\,r\notin A,s \in A} \l^A\uk
\]
of $\hK_{rs}$ and $\hK_{sr}$, we note that
\begin{align*}
   &\sum_{A\colon\,r\in A,s \notin A} \MN(j;p_1,\ldots,p_R)\{\mm_{Aj}\}\\
     &{}\qquad= (1-p_s)^j\{1 - (1-p_r/(1-p_s))^j\} \\
     &{}\qquad= (1-p_s)^j - (1-p_r-p_s)^j,
\end{align*}
the probability under the multinomial scheme that the $r$-th cell is non-empty
but the $s$-th cell is empty.  Thus
\[
   \l_{rs} \Eq \sji \frac{\th_k}j\,\{(1-p_s)^j - (1-p_r-p_s)^j\} \Eq \th_k \log((p_r+p_s)/p_s),
\]
and $\l_{sr} = \th_k \log((p_r+p_s)/p_r)$.  Then, because $\hK_{rs}$ and $\hK_{sr}$ are
independent and Poisson distributed,
\[
  \ex\{(\hK_{rs}-\hK_{sr})^2\} \Eq (\l_{rs} - \l_{sr})^2 + \l_{rs} + \l_{sr}.
\]
This yields the formula
\eq\label{mean-stat}
  \m \Def \frac1{R(R-1)}\, \sum_{1\le r < s \le R} \Blb \th_k^2\{\log(p_r/p_s)\}^2
     + \th_k \log\Bl \frac{(p_r+p_s)^2}{p_rp_s}\Br \Brb.
\en
In particular, if $p_r = 1/R$ for $1\le r\le R$, then
$\m = \th_k \log 2$, agreeing with the observation of \citet{wje07}\index{Ewens, W. J.}\index{RoyChoudhury, A.}\index{Lewontin, R. C.}\index{Wiuf, C.}
in the case $R=2$.\index{cancer|)}

\section{Conclusion}

Our paper is about \Index{ancestral inference} (albeit in a
somatic \Index{cell} setting rather than the typical population
genetics\index{mathematical genetics} one) and Poisson approximation\index{Poisson, S. D.!Poisson approximation|)}. John Kingman has made fundamental
and far-reaching contributions to both areas\index{Kingman, J. F. C.!influence}.
It therefore gives us great pleasure to dedicate it to John on his birthday.

\paragraph{Acknowledgements}
ST  acknowledges the support of the University of Cambridge, Cancer Research UK and Hutchison Whampoa Limited. ADB was supported in part by Schweizer National\-fonds Projekt 
Nr.\ 20--117625/1. 

\bibliographystyle{cambridgeauthordateCMG}
\bibliography{adhoc}

\end{document}